\documentclass[aps,prl,twocolumn,groupedaddress]{revtex4}

\newcommand{\PL}[3]{Phys. Lett. {\bf #1}, #2 (#3)}
\newcommand{\PRL}[3]{Phys. Rev. Lett. {\bf #1}, #2 (#3)}
\newcommand{\NP}[3]{Nucl. Phys. {\bf #1}, #2 (#3)}
\newcommand{\PR}[3]{Phys. Rev. {\bf #1}, #2 (#3)}
\newcommand{\PTP}[3]{Prog. Theor. Phys. {\bf #1}, #2 (#3)}

\usepackage{graphicx}

\begin{document}


\title{High-density \={K} nuclear systems with isovector deformation}


\author{Akinobu Dot$\acute{\rm e}$}
\affiliation{Institute of Particle and Nuclear Studies, KEK, Tsukuba, Ibaraki
305-0801, Japan}

\author{Hisashi Horiuchi}
\affiliation{Department of Physics, Kyoto University, Kyoto 606-8502, Japan}

\author{Yoshinori Akaishi}
\affiliation{Institute of Particle and Nuclear Studies, KEK, Tsukuba, Ibaraki
305-0801, Japan}

\author{Toshimitsu Yamazaki}
\affiliation{RI Beam Science Laboratory, RIKEN, Wako,
Saitama 351-0198, Japan}


\date{\today}

\begin{abstract}
Using a phenomenological
\={K}N potential which reproduces $\Lambda$(1405) as an
$I=0$ bound state of \={K}N, we investigated deeply bound kaonic
nuclei, ppnK$^-$ and $^8$BeK$^-$,
with the method of Antisymmetrized Molecular Dynamics.
Our calculations show that strongly bound kaonic nuclear systems with
unusual exotic structures  are formed around the K$^-$, which attracts the
surrounding nucleons to an extremely high-density assembly and induces a
proton-neutron separation, ``isovector deformation''.
\end{abstract}


\maketitle



Recently, exotic nuclear systems involving a $\bar{\rm K}$ (K$^-$ or 
\={K}$^0$) as a constituent
have been studied theoretically by Akaishi and Yamazaki
\cite{Akaishi-Yamazaki} (to be abbreviated as AY), who constructed  \={K}N
interactions phenomenologically so as to reproduce  low energy \={K}N
scattering data \cite{Martin},   kaonic hydrogen atom data \cite{Iwasaki}
and  the
binding energy and decay width of $\Lambda$(1405), which is asserted to be
an $I=0$ quasi-bound state of \={K}N. These interactions are  characterized
by a strongly attractive $I=0$ part.
Such properties of the AY \={K}N interactions are consistent with those
obtained by  using a chiral SU(3) effective Lagrangian by Weise {\it et al.}
\cite{Weise:96}.
With these interactions AY calculated the bound systems of ppnK$^-$,
ppnnK$^-$ and $^8$BeK$^-$, showing the following characteristics:  i) The $I=0$
\={K}N interaction is strong enough to shrink the nucleus against the
nuclear imcompressibility. ii)  The binding energies are extremely large due to
the strong $I=0$ \={K}N interaction helped by the nuclear shrinkage effect
so that the bound states lie below the threshold of the main decay channel to
$\Sigma\pi$, thus inferring the presence of quasi-stable discrete bound
states (width ($\Gamma_{\rm K}$) $<$ binding energy ($B_{\rm K}$)). 
Furthermore, Yamazaki
and Akaishi \cite{YA} have shown that the (K$^-,\pi^-$) reaction as a 
source of $\Lambda^*$ can produce
  various exotic \={K} bound states in non-existing
proton-rich nuclei,  such as p$^2$ ($^2$He), p$^3$ ($^3$Li), p$^3$n
($^4$Li), p$^4$n$^2$ ($^6$Be) and p$^5$n$^4$ ($^9$B). Since the presence of
a K$^-$ in nuclei is expected to cause a drastic change in the nuclear
systems, it is vitally important to perform theoretical studies of such 
exotic systems without any constraint from the existing
common knowledge of nuclear physics (shell and cluster structure, nuclear
shape, nuclear density, proton-neutron distributions, etc.). Thus, we
started a series of calculations, employing the method of Antisymmetrized
Molecular Dynamics (AMD) \cite{AMD:Enyo,AMD:Dote}, based on the AY \={K}N
interaction. The present paper reports the first results on ppnK$^-$ and
$^8$BeK$^-$.
 
The AMD method has a great advantage: it treats ``a nucleus + K$^-$" as a full
($A + 1$)-body
system without any assumption concerning {\it a priori} structure (shell or 
clucter).
Actually, AMD has succeeded to explain a variety of exotic shapes
and clusters in light unstable nuclei \cite{AMD:Enyo}.
If a \={K} nuclear system prefers a
structure where the proton distribution differs from the neutron distribution,
such a structure will be dynamically formed in the AMD treatment.

In the usual AMD, a single nucleon wave function, $| \varphi_i \rangle$,
is represented by a single Gaussian wave packet.
In the present study, however,
it is given as a superposition of several
Gaussian wave packets as follows, in order to describe the system more
precisely,  and the \={K} is represented
in the same way:


\begin{figure*}
\begin{minipage}[t]{5cm}
   \includegraphics[width=1.00\textwidth]{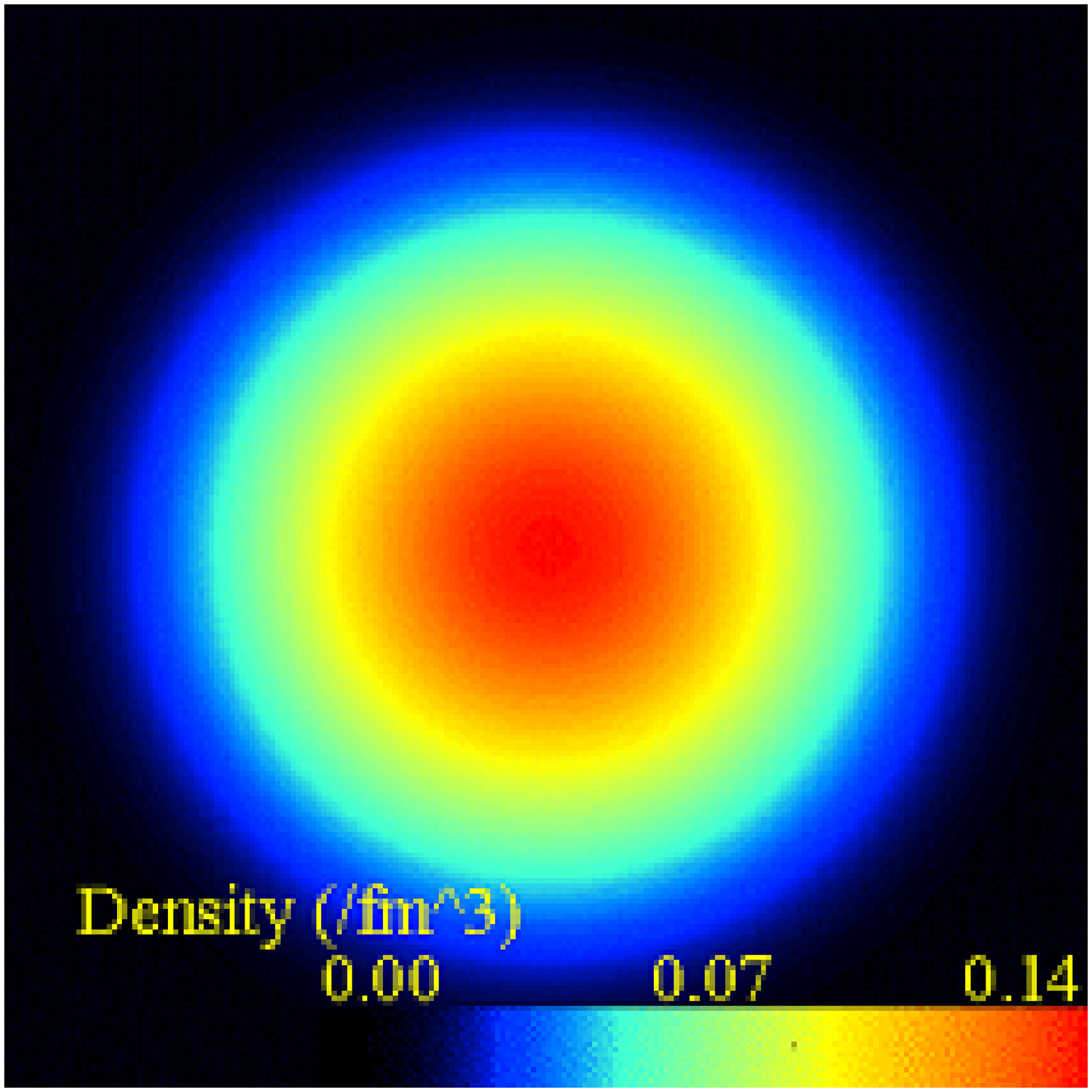}%

(a) $^3$He
\end{minipage}
\hspace{0.3cm}
\begin{minipage}[t]{5cm}
   \includegraphics[width=1.00\textwidth]{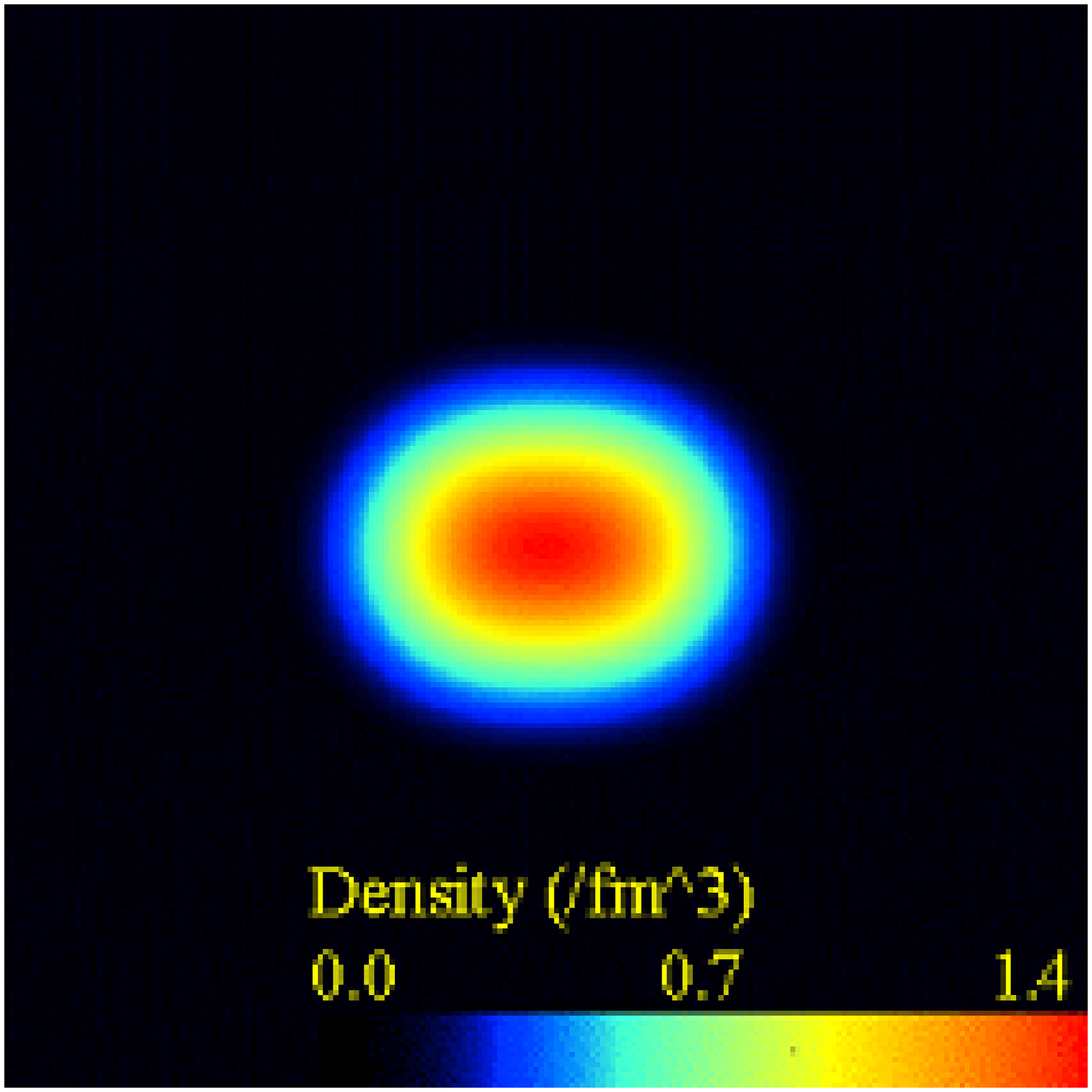}%

(b) $^3$HeK$^-$
\end{minipage}

\vspace{0.2cm}

\begin{minipage}[t]{3cm}
   \includegraphics[width=1.00\textwidth]{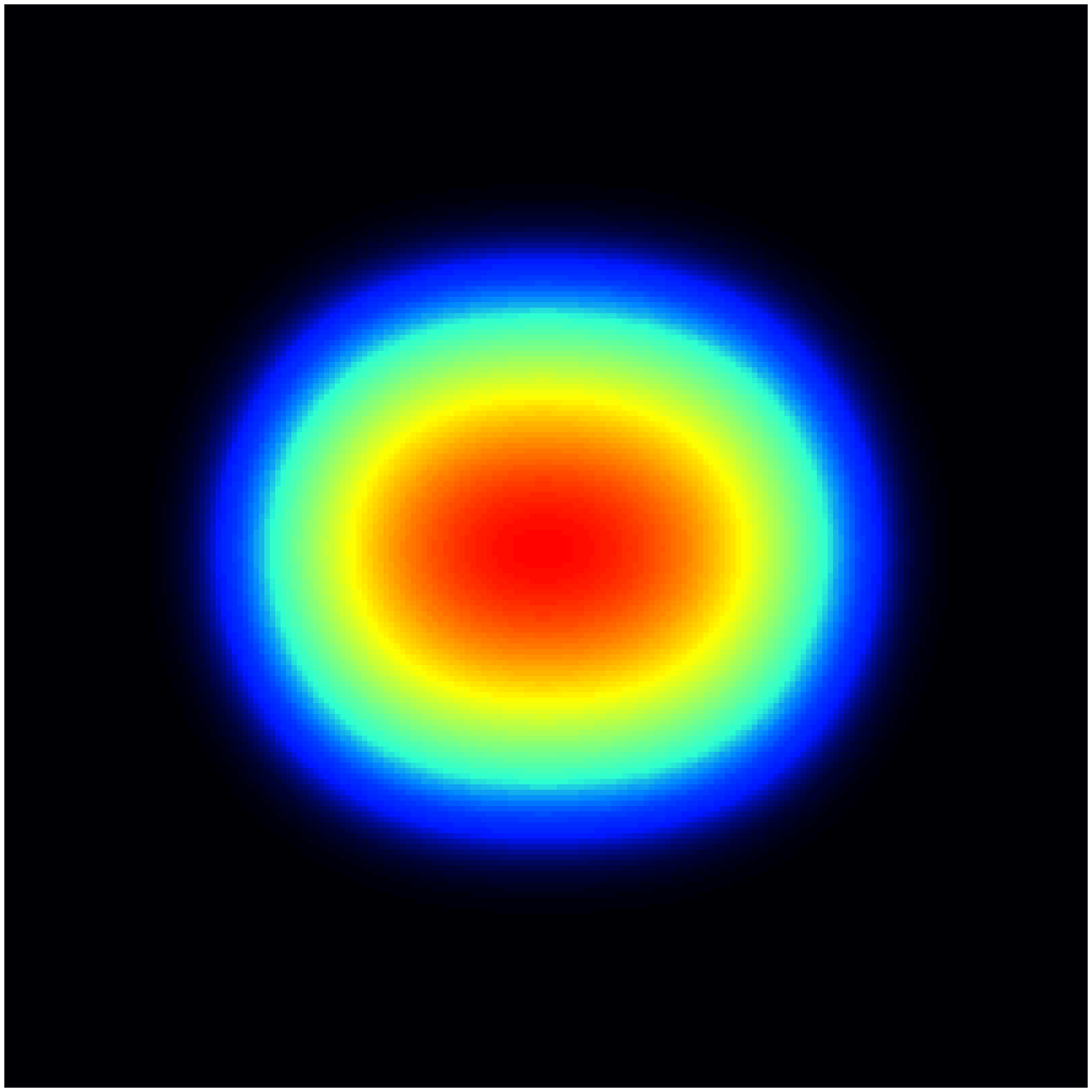}%

(c) Proton
\end{minipage}
\hspace{0.1cm}
\begin{minipage}[t]{3cm}
   \includegraphics[width=1.00\textwidth]{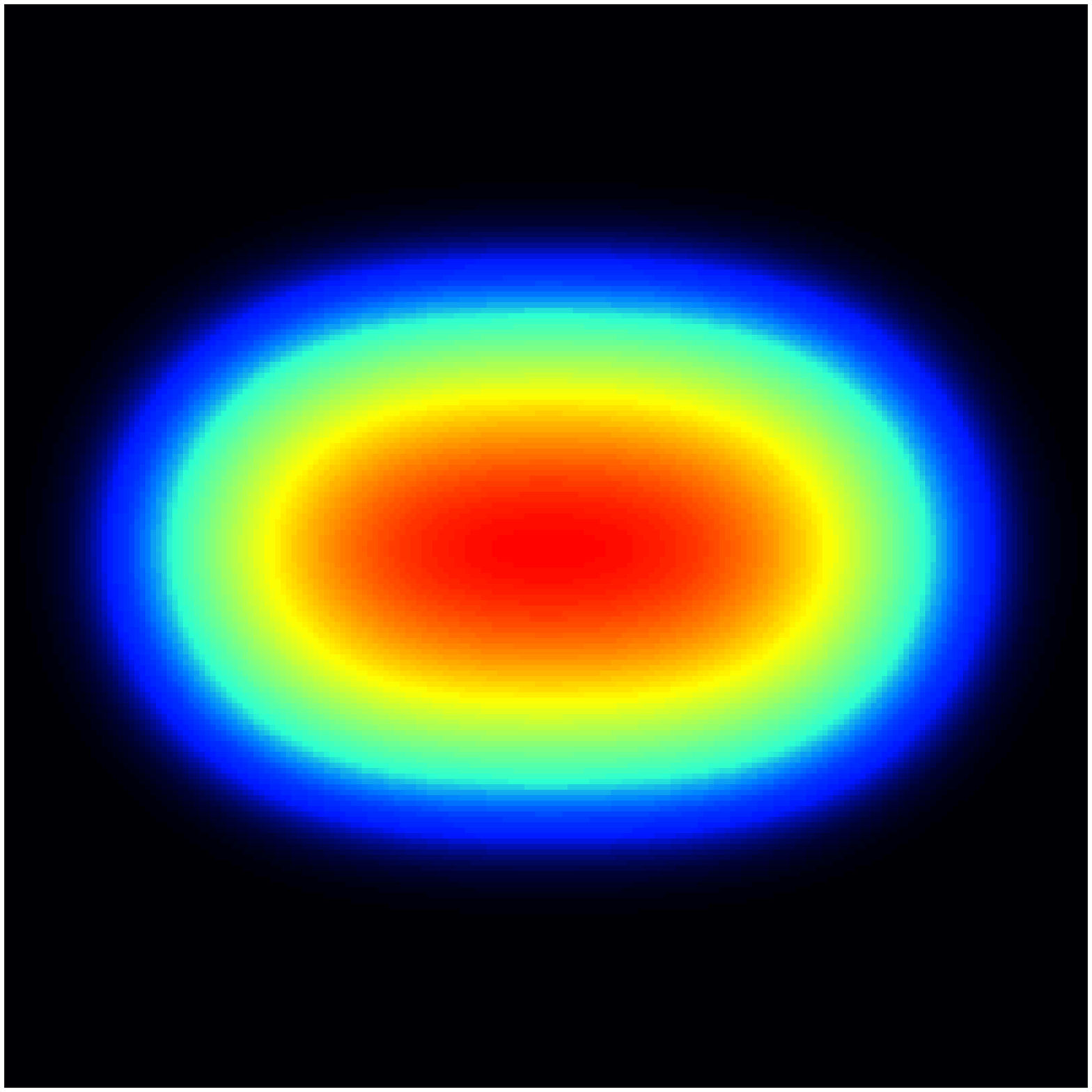}%

(d) Neutron
\end{minipage}
\hspace{0.1cm}
\begin{minipage}[t]{3cm}
   \includegraphics[width=1.00\textwidth]{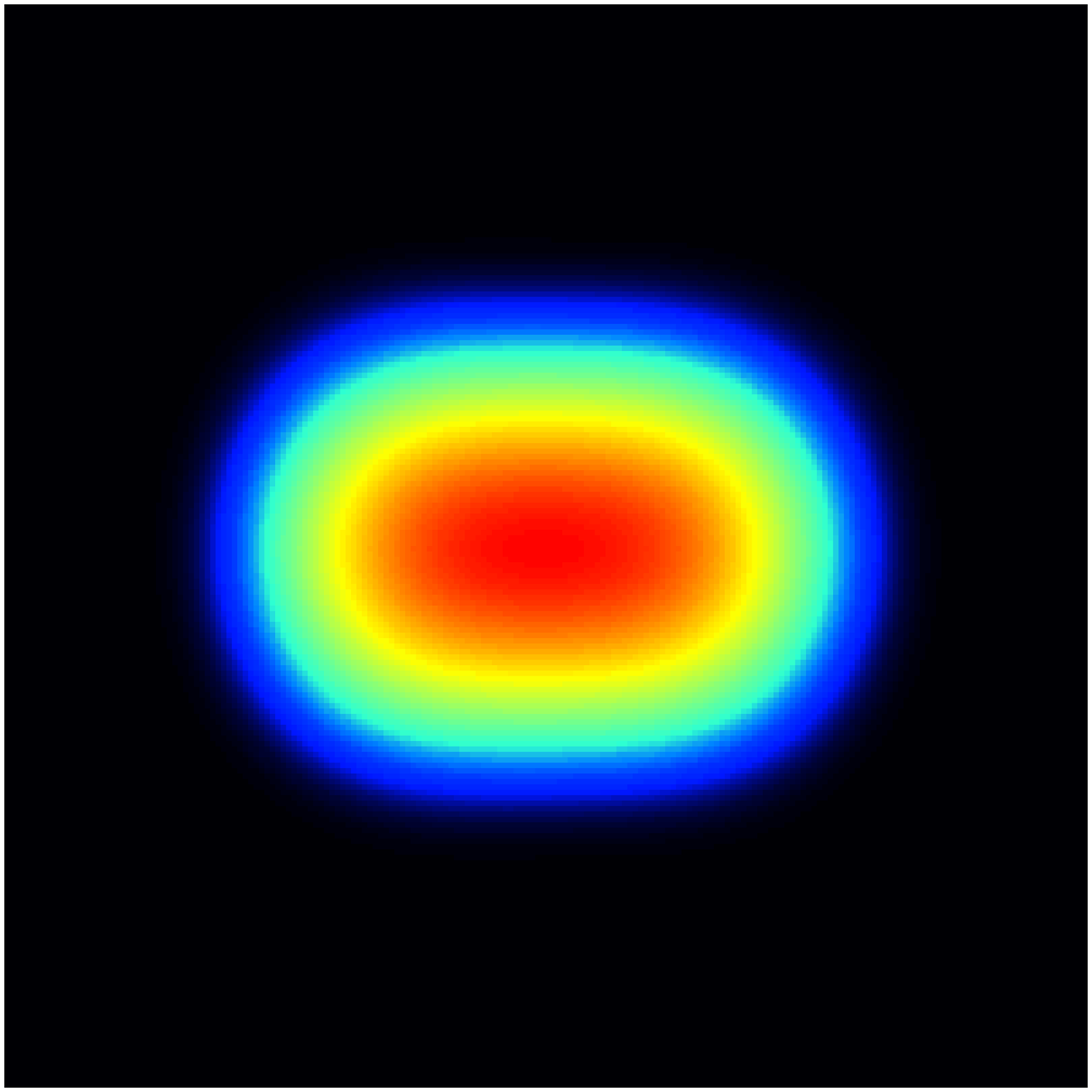}%

(e) K$^-$
\end{minipage}

   \caption{\label{}Calculated density contours of ppnK$^-$.
  Comparison between (a) usual $^3$He and
(b) $^3$HeK$^-$ is shown in the size of 5 by 5 fm.
Individual contributions of (c) proton, (d) neutron and (e) K$^-$ are given
in the size of 3 by 3 fm. }
\end{figure*}

\begin{eqnarray}
&  |\varphi_i \rangle&   = 
\sum_{\alpha=1}^{n_{\rm N}} C^i_\alpha
\exp \left[-\nu \left({\bf r}-\frac{{\bf Z}^i_\alpha}{\sqrt{\nu}}
\right)^2 \right]\;
|\sigma_i \tau_i \rangle , \\
&  |\varphi_{\rm K} \rangle&   =  
\sum_{\alpha=1}^{n_{\rm K}} C^{\rm K}_\alpha
\exp \left[-\nu \left({\bf r}-\frac{{\bf Z}^{\rm K}_\alpha}{\sqrt{\nu}}
\right)^2 \right]\;
|  \tau_{\rm K} \rangle ,
\end{eqnarray}
where $|\sigma_i \tau_i \rangle$ is a spin-isospin wave function,
which is $| {\rm p} \uparrow \rangle, | {\rm p} \downarrow \rangle
, | {\rm n} \uparrow \rangle$, or $| {\rm n} \downarrow
\rangle$;
$| \tau_{\rm K} \rangle$ indicates
the isospin wave function of \={K}.
We use two Gaussian wave packets for a single nucleon ($n_{\rm N}=2$)
and five for a \={K} ($n_{\rm K}=5$).
We must antisymmetrize the
   nucleon wave functions, which are then combined with a \={K} wave function as
\begin{eqnarray}
|\Phi  \rangle & = & \det[|\varphi_i(j)\rangle] \otimes
|\varphi_{\rm K} \rangle.
\end{eqnarray}
Finally, we project the intrinsic wave function $|\Phi \rangle$
onto the eigen-state of parity:
$|\Phi^\pm\rangle = \; |\Phi \rangle\pm {\cal P}|\Phi \rangle .$
We utilize $|\Phi^\pm\rangle$ as a trial function, which
contains complex variational parameters \{$C^i_\alpha,
{\bf Z}^i_\alpha$, $C^{\rm K}_\alpha,
{\bf Z}^{\rm K}_\alpha$\} and a real variational parameter, $\nu$.
They are determined by the frictional cooling equation
with constraint \cite{AMD:Dote}.


The Hamiltonian,
\begin{eqnarray}
\hat{H}  & = &  \hat{T} + \hat{V}_{\rm NN}  + \hat{V}_{\rm KN} + 
\hat{V}_{\rm C} - \hat{T}_{\rm G},
\end{eqnarray}
is composed of the kinetic energy, $\hat{T}$,
two-body NN interaction, $\hat{V}_{\rm NN}$, \={K}N interaction, 
$\hat{V}_{\rm KN}$,
and Coulomb force, $\hat{V}_{\rm C}$.
The center-of-mass motion energy, $\hat{T}_{\rm G}$, is subtracted.
In the kinetic energy and the c.m. motion energy,
we treat the mass difference between a nucleon and K$^-$ exactly.
Since the system is likely to be
extremely dense due to the strongly attractive \={K}N interaction,
we can use no existing effective interactions. Therefore, we use the 
Tamagaki potential (OPEG) \cite{Tamagaki} as a bare NN
interaction, and the
AY \={K}N interaction as a bare \={K}N interaction, and
  construct the effective NN-central force and the \={K}N force
with the G-matrix method \cite{Akaishi-Yamazaki}.
Since the Tamagaki potential
reproduces the NN phase shifts up to 660 MeV in laboratory energy
\cite{Tamagaki},
we expect that it is applicable to extremely high-density states of
relevant concern.


\begin{table}
\caption{\label{}
Summary of the present calculations.
$\rho (0)$: nucleon density at the center of the system. $R_{\rm rms}$:
  root-mean-square radius of the nucleon system.
$\nu$:   width parameter of a Gaussian wave packet used in the calculation.
$\beta$: deformation parameter for the nucleon system. ppnK$^-$ $^\dagger$ 
and $^8$BeK$^-$ $^\dagger$: AY's results.
}
\begin{ruledtabular}
\begin{tabular}{l|cccccc}
     & B.E.  &  $\Gamma_{\rm K}$ & $\rho (0)$ & $R_{\rm rms}$ &  $\nu$ & 
$\beta$ \\
     & [MeV] & [MeV]      & [fm$^{-3}$] & [fm] & [fm$^{-2}$] &       \\
\hline
$^3$He &   $5.95$         &   ---         &   0.14          & 1.59 &  0.20
& 0.0     \\
ppnK$^-$ & $113$       &  24        &   1.39          & 0.72 &  1.12
& 0.19 \\
ppnK$^-$ $^\dagger$ & $116$ & 20           &   1.10              & 0.97 &
&      \\
\hline
$^8$Be &  $48.72$         &   ---         &   0.10          & 2.46 &  0.205
& 0.63 \\
$^8$BeK$^-$ & $ 159$   &  43        &   0.76          & 1.42 &  0.52
& 0.55 \\
$^8$BeK$^-$ $^\dagger$ & $168$ &  38         & $\sim$ 0.85 &  &
&      \\
\end{tabular}
\end{ruledtabular}
\end{table}

We have calculated  systems of ppnK$^-$ and $^8$BeK$^-$. These systems
were previously treated with a  Brueckner-Hartree-Fock method by AY
\cite{Akaishi-Yamazaki}.
We summarize our calculations for ppnK$^-$ and $^8$BeK$^-$ in Table I.
We also show the AMD results of the normal $^3$He and $^8$Be for a comparison.
$^3$He was calculated with the Volkov No.1 force \cite{Volkov}
and $^8$Be with the MV1 Case-3 force \cite{MV1}. However, we did
not perform
an angular momentum projection for simplicity.
$\Gamma_{\rm K}$ is the width for decaying to $\Lambda\pi$ and $\Sigma\pi$, and
   is evaluated by calculating the expectation value
of the imaginary potential
contained in the effective  AY \={K}N interaction with the wave function
obtained by the AMD calculation. ``ppnK$^-$ $^\dagger$'' and ``$^8$BeK$^-$
$^\dagger$''
are the results of AY.

\begin{figure*}
\begin{minipage}[t]{0.3\textwidth}
   \includegraphics[width=1.00\textwidth]{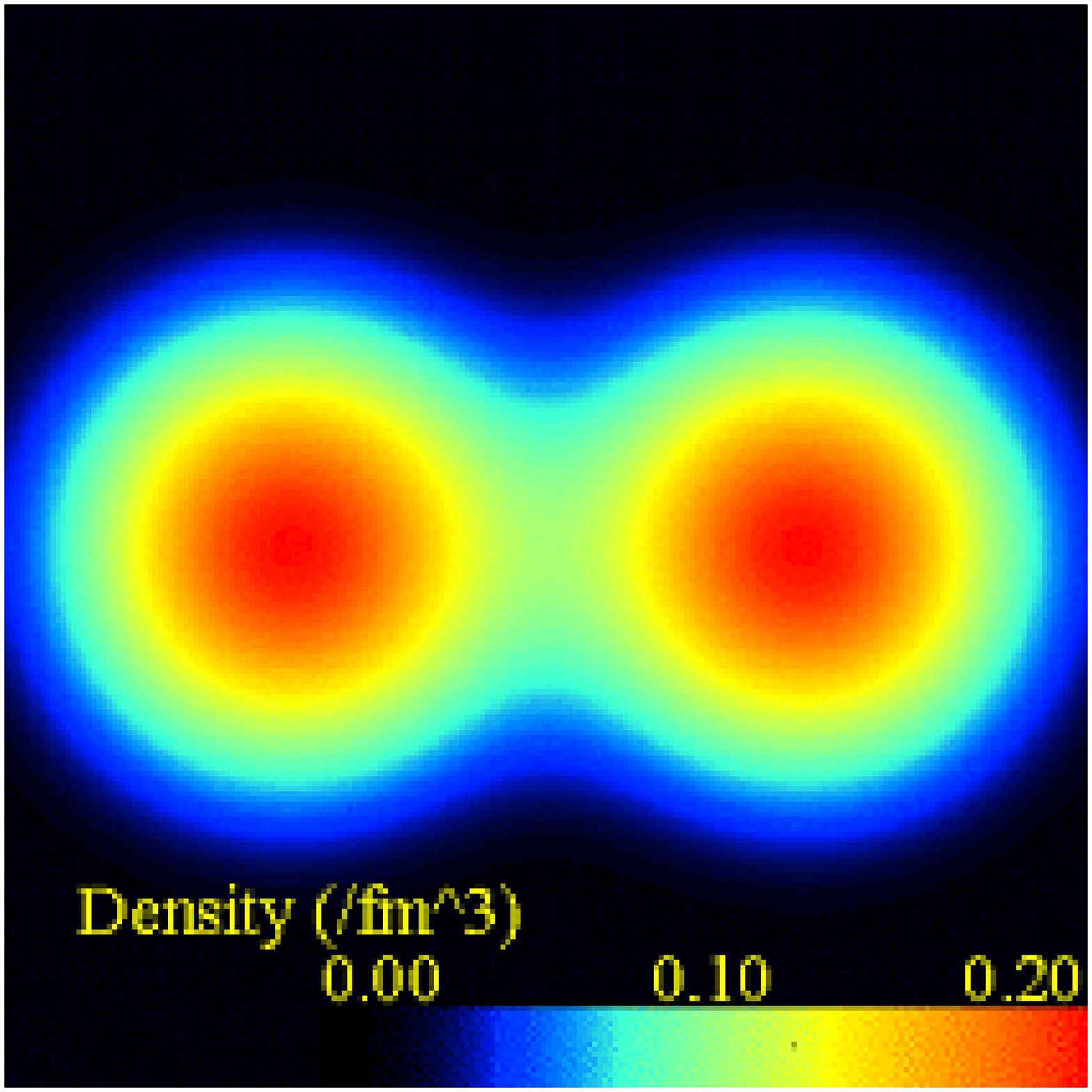}%

(a) $^8$Be
\end{minipage}
\hspace{0.3cm}
\begin{minipage}[t]{0.303\textwidth}
   \includegraphics[width=1.00\textwidth]{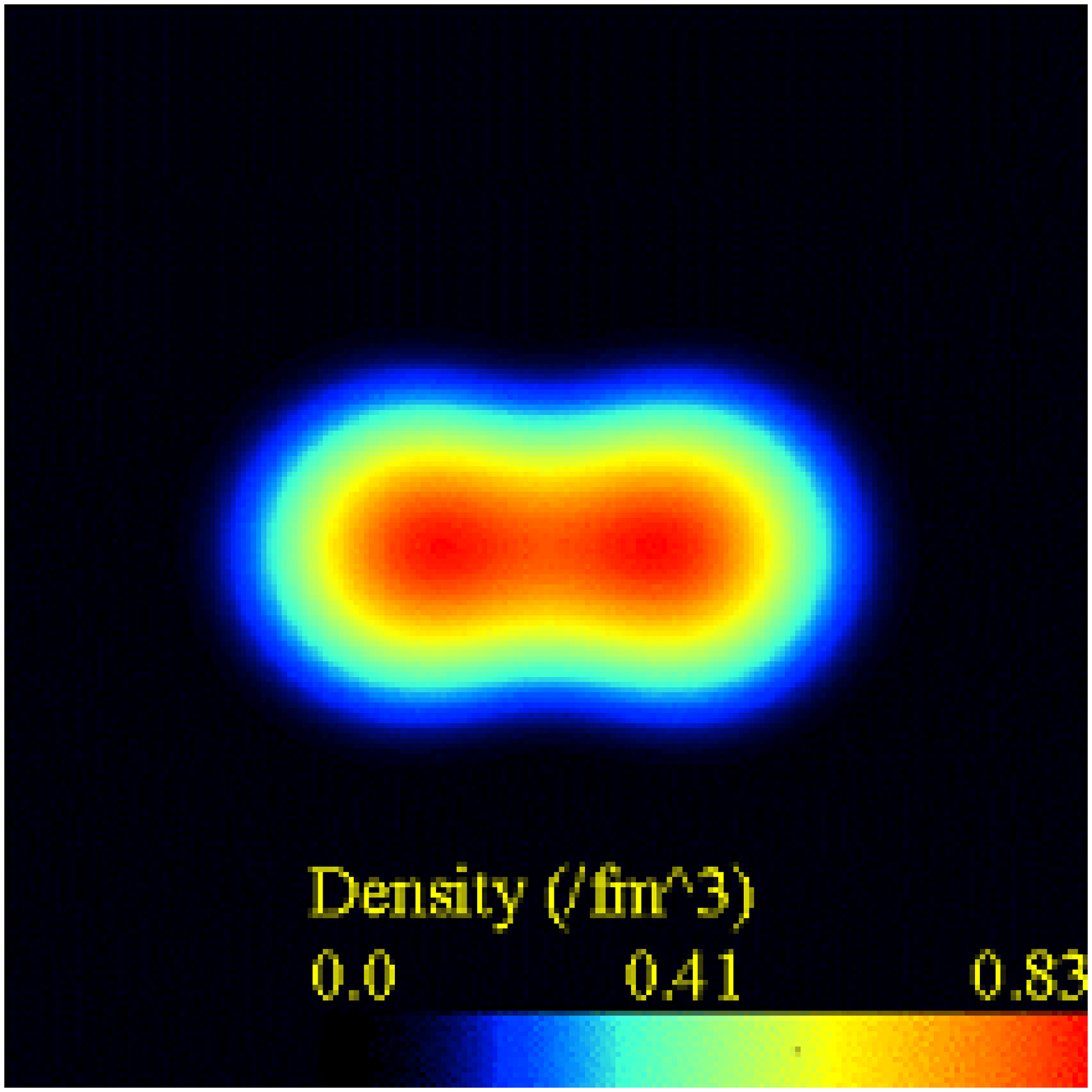}%

(b) $^8$BeK$^-$
\end{minipage}

\vspace{0.2cm}

\begin{minipage}[t]{3.1cm}
   \includegraphics[width=1.00\textwidth]{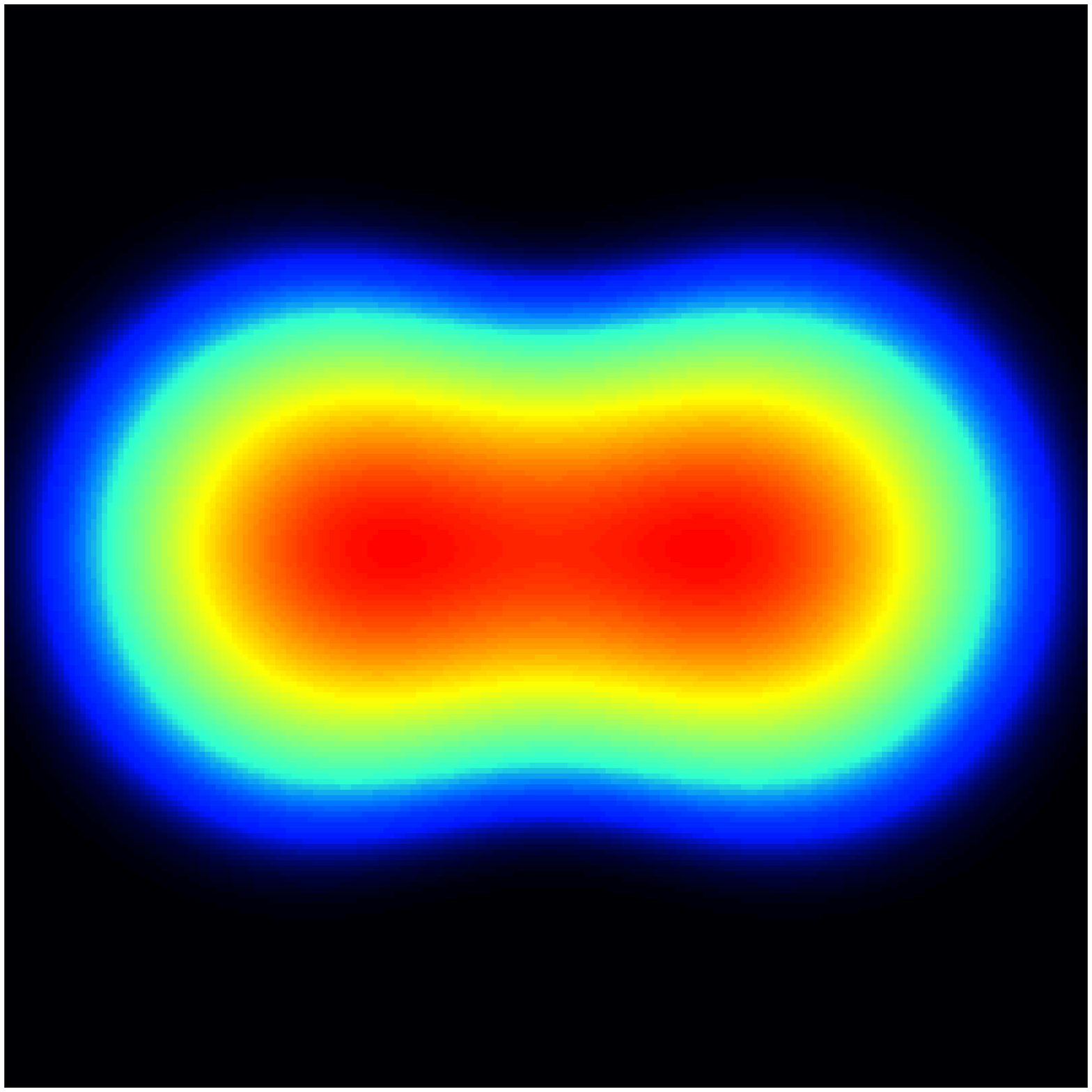}%

(c) Proton
\end{minipage}
\hspace{0.1cm}
\begin{minipage}[t]{3.1cm}
   \includegraphics[width=1.00\textwidth]{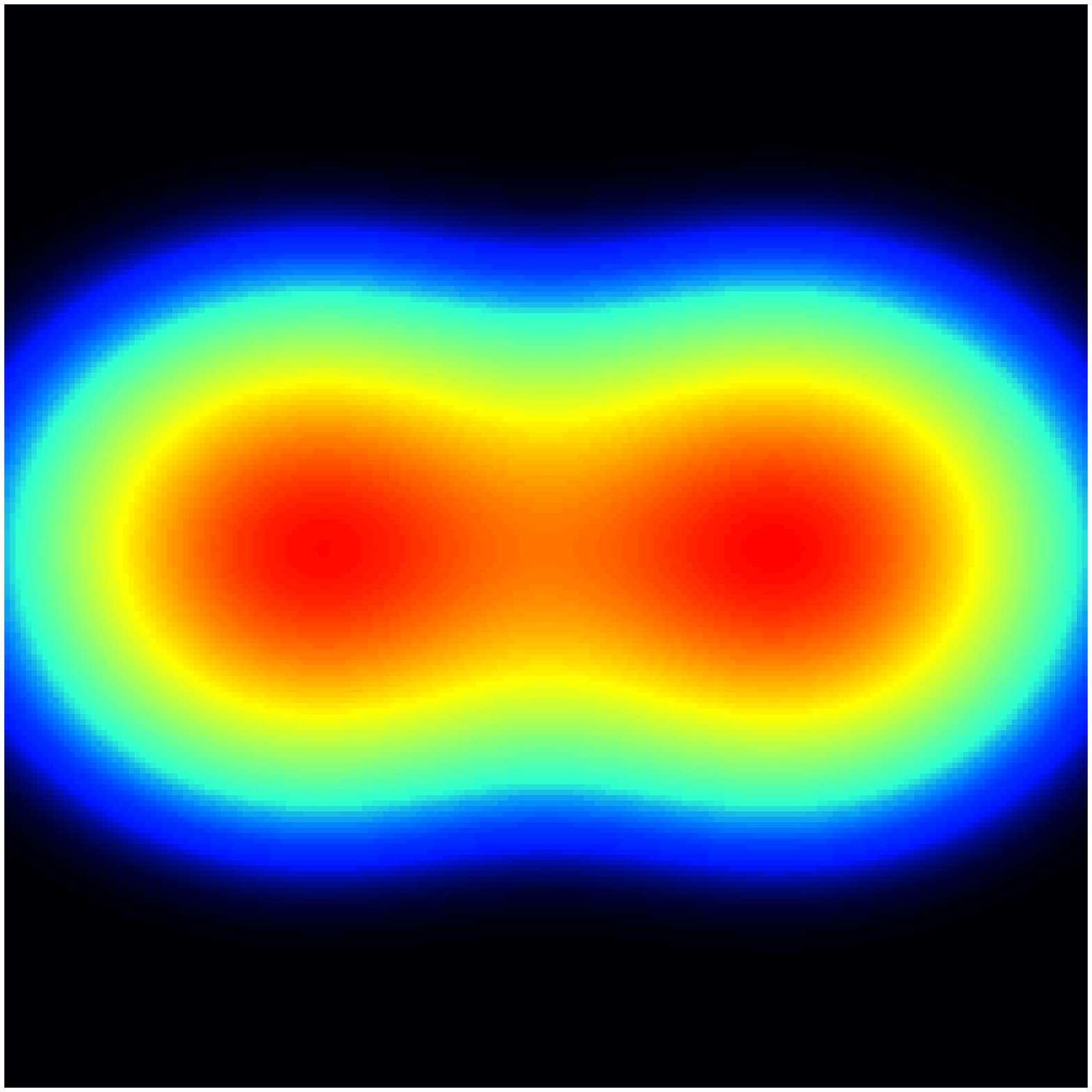}%

(d) Neutron
\end{minipage}
\hspace{0.1cm}
\begin{minipage}[t]{3.1cm}
   \includegraphics[width=1.00\textwidth]{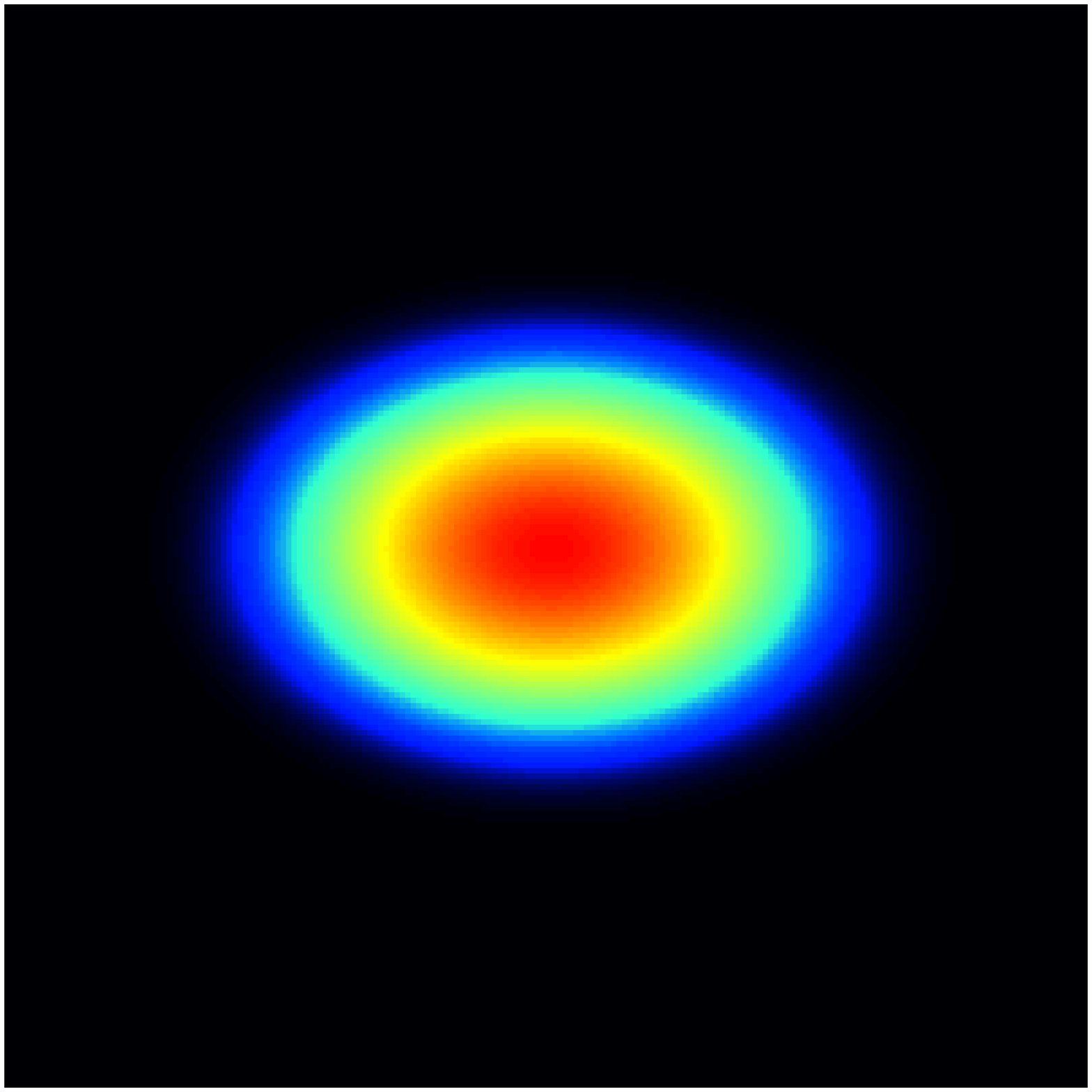}%

(e) K$^-$
\end{minipage}
   \caption{\label{}Calculated density contours of $^8$BeK$^-$.
  Comparison of the density distributions of (a) usual $^8$Be and
(b) $^8$BeK$^-$ is shown in the size of 7 by 7 fm.  Individual 
contributions of (c) proton, (d) neutron and (e) K$^-$ are given in the 
size of 4 by 4 fm.
}
\end{figure*}


Our result shows that the ppnK$^-$ ($T=0$) system is bound by 113 MeV 
($B_{\rm K}$ =105 MeV
below the $^3$He+K$^-$ threshold). Since the obtained state lies under
the $\Sigma \pi$ threshold ($-80$ MeV), the $\Sigma \pi$ channel,
which is the main decay-channel, is closed and the decay width, 
$\Gamma_{\rm K}$,
becomes narrow to be 24 MeV. The present result is very similar to the
AY prediction: $B_{\rm K} =108$ MeV and $\Gamma_{\rm K} = 20$ MeV.  We have 
not considered
the decay width from the non-mesonic decay
(\={K}NN $\rightarrow$ $\Lambda$N/$\Sigma$N), but according to
AY it is estimated to be about 12 MeV \cite{Akaishi-Yamazaki}.
The width of ppnK$^-$ remains still narrower than that of $\Lambda(1405)$,
even when the non-mesonic decay is taken into account.

Surprisingly,
the central density (``uncorrelated density'') of the system amounts
to 8.2-times the normal density due to the shrinkage effect. Figs. 1(a) and 
(b) show a comparison between $^3$He and $^3$HeK$^-$.  In order to
see how the bound \={K} changes the nucleus in more detail we show the 
calculated density
distributions of the constituents in Figs.1(c)-(e).
Apparently, the proton distribution is more compact than the
neutron distribution. This phenomenon is attributed to the property of
the \={K}N interaction.
Table II shows how protons and a neutron in ppnK$^-$ contribute to
the kinetic energy and the expectation value of the \={K}N interaction,
and also to each root-mean-square radius.
This table together with Fig.1 can be interpreted as follows.
Since the K$^-$p interaction is much stronger than the K$^-$n one,
the protons distribute compactly near the K$^-$.
Though the kinetic energy of the protons increases, the
total energy decreases due to the strongly attractive K$^-$p interaction.
On the other hand, the neutron is not favoured by shrinking, since the 
K$^-$n interaction is not attractive enough to compensate for the increase 
of the kinetic energy.

\begin{table}
\caption{\label{}
Contributions of proton and neutron in ppnK$^-$.
$\langle T \rangle$: kineic energy per nucleon.
$\langle V_{KN} \rangle$:  expectation value of the \={K}N interaction per 
nucleon.
$R_{\rm rms}$: root-mean-square radii of the
proton and neutron distributions.
}
\begin{ruledtabular}
\begin{tabular}{l|ccc}
          & $\langle T \rangle$ & $\langle V_{\rm KN} \rangle$ & $R_{\rm 
rms}$  \\
          & [MeV/u]  & [MeV/u]  & [fm]  \\
\hline
proton  & 69.4 & $-170.2$     & 0.70  \\
neutron & 59.0 & $-42.4$      & 0.75  \\
\end{tabular}
\end{ruledtabular}
\end{table}


Now we proceed to $^8$BeK$^-$. As expected, $^8$Be is extremely shrunk due 
to the presence of a K$^-$. The K$^-$ is bound by 159 MeV ($B_{\rm K}$ = 
104 MeV with respect to  $^8$Be).
The binding energy of K$^-$ and the decay width
obtained by our calculation are very similar to those
   by AY, who employed the $\alpha$$\alpha$K$^-$ model
\cite{Akaishi-Yamazaki}.
We compare the density distribution of $^8$BeK$^-$
with that of the normal $^8$Be in Figs.2(a) and (b).
$^8$BeK$^-$ becomes much more compact than $^8$Be, and attains a high 
density,  4.5-times
the normal nuclear density.
As shown in Table I, the root-mean-square radius decreases dramatically
from 2.46 fm to 1.42 fm. Nevertheless,
   an $\alpha$-cluster like structure still remains in $^8$BeK$^-$.
We can see that the relative distance between the $\alpha$ clusters becomes
short and that the $\alpha$ cluster, itself, in $^8$Be is also shrunk.
The shrinkage of each $\alpha$ cluster is indicated by
that of the Gaussian wave packet, as shown in Table I.
The width parameter, $\nu$, is changed from 0.205 fm$^{-2}$ for the normal 
$^8$Be
to 0.52 fm$^{-2}$ for $^8$BeK$^-$. The deformation parameter, $\beta$, 
becomes slightly  smaller.

Figs.2(c)-(e) display contour mapping of the calculated  density 
distributions of
$^8$BeK$^-$
in more detail.
Although the two-$\alpha$ like structure, as shown in (b), persists,
   it is found that the
proton  distribution (c) is significantly different from the neutron 
distribution (d).
As in the case of  ppnK$^-$, the protons feel
a strongly attractive force from the K$^-$, which stays in the center of
the total system (Fig.2(e)), and
   are attracted nearer to the K$^-$ than the neutrons.
Namely, the proton distribution is separated from the neutron distribution
in spite of $N=Z$. A more quantitative presentation of the proton and
neutron density distributions is given in Fig.3. The \={K} produces a new
type of neutron halo; it is peripherally distributed, yet with quite a high 
density.
We name such a new type of deformation as an ``isovector
deformation''. It should be noted that AMD enables us to predict such an
isovector deformation
without any assumption about resultant structure. Such a deformation cannot 
be seen
in the calculation
by AY because they assumed the existence of $\alpha$ clusters.

\begin{figure}
   \includegraphics[width=0.5\textwidth]{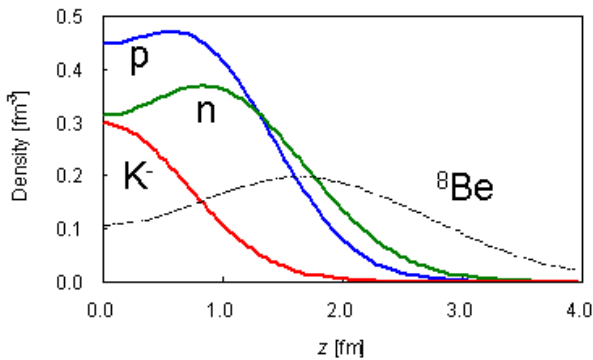}%
   \caption{\label{} Calculated density distributions of proton, neutron and 
K$^-$ in $^8$BeK$^-$ along
the horizontal axis in Fig.2.
For comparison the density distribution of $^8$Be
is also displayed as ``$^8$Be''.}
\end{figure}


In summary, we have shown from our AMD calculations that a K$^-$ forms a
deeply bound nuclear state by producing an {\it extremely high-density}
nuclear system with {\it isovector deformation}.  No such nuclear systems
at ``zero temperature and high density" have been created so far. Whereas an
experiment to identify $^3$HeK$^-$ from $^4$He(stopped K$^-$, n)
\cite{Akaishi-Yamazaki,Iwasaki:01} is under way at KEK, a new possibility
to make use of (K$^-, \pi^-$) reactions to populate proton-rich systems has
been proposed \cite{YA}. We are extending our AMD
calculations to such more exotic systems, as AMD is the only possible
method to investigate their dynamics without any {\it a priori}  structural 
assumption. If a deeply bound kaonic nucleus is confirmed experimentally,
new developments of nuclear physics will follow. This new field of 
``bound-\={K} nuclear
spectroscopy" will  create a new {\it dense and cold} environment, in 
contrast to  hot and dense states expected to be realized by heavy-ion 
collisions. Restoration of chiral symmetry in a dense nuclear medium might
be observed for the \={K}N system
in the same way as in deeply bound pionic states
\cite{Weise:00,Kienle,pi_205Pb}. The bound-\={K} nuclear
spectroscopy will yield
  unique information about  a transition from the hadron phase to a hitherto 
untouched
``quark phase". It will  also provide  experimental grounds for possible kaon
condensation and strange matter formation \cite{Brown}.
\\

One of the authors (A.D.) thanks Dr. M. Kimura for giving him
the graphic tools he made. The present work is supported by Grants-in-Aid
for Scientific Research of Monbukagakusho of Japan.

\bibliography{apssamp}

\end{document}